\documentstyle[12pt, a4]{article}

\input epsf

\begin{document}

\title{Inertial frame dragging and Mach's principle in 
General Relativity}
\author{
  A.R. Prasanna\thanks{prasanna@prl.ernet.in}\\
  Physical Research Laboratory\\
  Ahmedabad, India
  }
\maketitle

\begin{abstract}
We define a new parameter `cumulative drag index' for a 
particle in circular orbit in a stationary, axisymmetric 
gravitational field and study its behaviour in the two well 
known solutions of general relativity {\it viz.}, the Kerr
spacetime and the G\"odel spacetime, wherein the inertial 
frame dragging has an important role. As it shows similar 
behaviour for both co and counter rotating particles, it
may indeed be an indication of the influence of the faraway
universe on local physics and thus Machian.
\end{abstract}

\newpage

\section{Introduction}
It is indeed well known that `Mach's principle' which 
relates `inertia' with the influence of distant sources in 
the universe in its original formulation has been a topic 
of interesting and intense discussions for a long time, particularly 
in the context of general relativity. For a recent 
survey of activities in this field, the best source is the proceeding of 
the T\"ubingen conference (1993) on this topic (Barbour and Pfister 1995 
\cite{BabPfi95}), wherein as many as twenty different 
interpretations of the principle have been discussed. 
Subsequent to this meeting there have been some interesting arguments with
conflicting conclusions concerning the celeberated Lense--Thirring effect 
with Rindler \cite{Rin} claiming this to be anti--Machian, while Bondi and
Samuel \cite{BonSam} claim it is Machian. These discussions do indeed pose a 
far more important question as to the nature of inertia and how one should
define `inertial frames' or inertial forces.

General relativity which is the most successful and complete theory of 
classical gravity, tried to do away with the concept of force by describing
gravity as the curvature of spacetime geometry. However in certain
contexts it may be useful to reintroduce the concept of force within the 
framework of general relativity and understand how the spacetime curvature
influences the various parts of the acceleration acting on a test particle 
in a given spacetime. This was indeed made possible by a 3+1 conformal 
splitting of spacetime by Abramowicz et al.~\cite{Abr_etal} and it has 
yielded several interesting new insights into the particle motion in 
curved spacetime.  In the context of Machian definition of inertia
one considers the induced effects of rotation in flatspace (Pfister and Braun
1985 \cite{PfiBra}) and the entire definition depends only on the relative 
rotation, in the absence of which both Coriolis, and the centrifugal 
accelerations are zero. On the other hand, in a curved spacetime, the 
contribution of curvature in the definition of inertial forces would change the
situation and could give a better analysis of the effects due to distant 
sources on local
physics. In fact the work due to Abramowicz and coworkers regarding the 
Newtonian forces in general relativity lead to some new discoveries like
centrifugal reversal, reversal of Rayleigh criterion (Abramowicz and
Prasanna 1990 \cite{AbrPra}), and the explanation of the ellipticity 
maximum of a rotating, stationary configuration (Abramowicz and
Miller \cite{AbrMil}). There have been certain other discussions, wherein
these new features are attributed to other reasons within the 
four--dim.~formalism without using the concept of inertial forces 
(de Felice \cite{deFe}, Barb\`es et al.~\cite{Bar_etal}). However in the 
context of Mach's principle and general relativity, it is imperative to 
introduce the `inertial forces' within the scope of general relativity and
possibly seperate the global and local effects.

Amongst the various effects of rotation, the most celebrated one is the
Lense--Thirring effect which arises due to `Coriolis force' coming from
the relative rotation.  It is amusing to note that this effect
has been considered to be both anti--Machian and Machian as mentioned above
thus leading to more confusion and leaving the conclusions  to 
the different interpretation of the principle. As the effect arises due to 
`dragging of inertial frames' it is best to define an index that 
incorporates the ratio of the relative acceleration to the total 
acceleration which could effectively subdue the contribution of the 
gravitational field and thereby allow one to get at the pure rotation effects.

In the following we attempt to consider this with the introduction of a new
dimensionless parameter called the `cumulative drag index' defined as the
ratio of the difference between the Coriolis and gravitational force to the
total force acting on a particle in circular orbit located at a distance 
where the centrifugal force acting on the particle is totally zero.
Generally, it is assumed that the Coriolis and the centrifugal forces arise
from the linear and the quadratic order of the angular velocity parameter
and in flatspace thus either both are zero or both are present. On the other
hand in general relativity the way we have introduced the `inertial
forces' using the optical reference geometry, it becomes apparent that
there do exist orbits along which the centrifugal force vanishes but
the Coriolis force can still be non zero. While in static spacetimes this 
orbit coincides with that of the unstable photon orbit (Prasanna 1991 
\cite{Pra}), in	stationary spacetime this orbit is different from the
photon orbits.

\section{Formalism}
In an axially symmetric stationary gravitational field represented by the
spacetime metric (with signature +2)
\begin{equation}
\label{axstat_metric}
ds^2 = g_{tt}\,dt^2 + 2 g_{t\varphi}\,dt\,d\varphi 
  + g_{\varphi\varphi}\,d\varphi^2 + g_{rr}\,dr^2 
  + g_{\theta\theta}\,d\theta^2
  \mbox{\raisebox{4ex}{}}
  \mbox{\raisebox{-4ex}{}}
\end{equation}
the four--acceleration $a^k$ acting on a particle in a circular orbit with
constant angular velocity $\Omega$, may be decomposed covariantly as given
by (Abramowicz et al.~1995)
\begin{equation}
\label{accel}
a_k = \nabla_k \Phi + \gamma^2 v \,(n^i \nabla_i \tau_k 
  + \tau^i \nabla_i n_k)
  + (\gamma v)^2 \tilde{\tau}^i\tilde{\nabla}_i \tilde{\tau}_k 
  \mbox{\raisebox{4ex}{}}
  \mbox{\raisebox{-4ex}{}}
\end{equation}
wherein the vector field $n^i$ corresponds to the locally non rotating 
observer (LNRF, Bardeen 1972 \cite{Bard}) expressed in terms of the 
timelike Killing vector $\eta^i$ and the spacelike Killing vector $\xi^i$:
\begin{equation}
\label{def_n_om}
n^i = e^\Phi (\eta^i + \omega\, \xi^i)\;, \quad 
\omega = -\langle\eta,\xi\rangle/\langle\xi,\xi\rangle \;.
  \mbox{\raisebox{4ex}{}}
  \mbox{\raisebox{-4ex}{}}
\end{equation}
Here $\Phi$ is the potential
\begin{equation}
\label{def_Phi}
\Phi = -\frac{1}{2}\ln[
   -\langle\eta,\eta\rangle
   -2\omega\langle\xi,\eta\rangle
   -\omega^2\langle\xi,\xi\rangle]\;,
   \mbox{\raisebox{4ex}{}}
  \mbox{\raisebox{-4ex}{}}
\end{equation}
$\tau^i$ is the unit vector orthogonal to $n^i$ along the circle, and $v$ 
the Lorentz speed related to the particle four--velocity $U^i$:
$U^i = \gamma (n^i + v\,\tau^i)$, $\gamma$ being the Lorentz factor 
$=1/\sqrt{1-v^2}$. For the circular orbit $U^i$ may also be decomposed as
$U^i = A (\eta^i + \Omega\,\xi^i)$ with $A$ being the redshift factor given
by:
\begin{equation}
\label{def_Aq}
A^2 = -[\langle\eta,\eta\rangle
   +2\Omega\langle\xi,\eta\rangle
   +\Omega^2\langle\xi,\xi\rangle]^{-1}\;.
  \mbox{\raisebox{4ex}{}}
  \mbox{\raisebox{-4ex}{}}
\end{equation}
In (\ref{accel}) the overhead tilde refers to the vectors defined in the
conformally projected three space (ACL 1988) with the positive definite 
metric
\begin{equation}
\label{def_hik}
h_{ik} = g_{ik} + n_i n_k\;, \quad
\tilde{\tau}^i = e^\Phi \tau^i
  \mbox{\raisebox{4ex}{}}
  \mbox{\raisebox{-4ex}{}}
\end{equation}
and $\tilde{\nabla}_i$ the covariant derivative with respect to the 
projected metric 
$\tilde{h}_{ik} = e^{2\Phi} h_{ik}$.
The three terms on the right hand side of (\ref{accel}) are respectively
identified as the gravitational, the Coriolis (Lense--Thirring), and the
centrifugal acceleration (Abramowicz 1993). It is easy to see that the 
particle speed $v$ is given by the relation
\begin{equation}
\label{def_v}
v\,\tau^i = e^\Phi (\Omega - \omega)\xi^i \;,
  \mbox{\raisebox{4ex}{}}
  \mbox{\raisebox{-4ex}{}}
\end{equation}
using which one can write explicitly the three forces acting on a particle
of rest mass $m_0$ (normalised) in the circular orbit with constant angular
velocity $\Omega$ as measured by the stationary observer at infinity, to be
\begin{eqnarray}
 & & \nonumber\\
\mbox{gravitational:}\quad 
G_i & = & \nabla_i\Phi = 
  -\frac{1}{2}\partial_i\left\{
  \ln[(g^2_{t\varphi} - g_{tt}\,g_{\varphi\varphi})/g_{\varphi\varphi}]
  \right\} 
  \label{def_Gi} \\
   & & \nonumber\\
\mbox{Coriolis:}\quad
(Co)_i & = & \gamma^2\,v\,n^j(\nabla_j\tau_i - \nabla_i\tau_j) 
   \nonumber \\
   & = & A^2 (\Omega - \omega)\sqrt{g_{\varphi\varphi}}
   \left\{
   \partial_i(\frac{g_{t\varphi}}{\sqrt{g_{\varphi\varphi}}}) 
   + \omega\,\partial_i\sqrt{g_{\varphi\varphi}}
   \right\}
   \label{def_Co} \\
    & & \nonumber
\end{eqnarray}
and centrifugal:
\begin{equation}
\label{def_Cf}
(Cf)_i = -\frac{(\gamma\,v)^2}{2}
   \tilde{\tau}^j\tilde{\nabla}_j\tilde{\tau}_i =
   -\frac{A^2 (\Omega - \omega)^2}{2} g_{\varphi\varphi}\partial_i
   \left\{
   \ln[g^2_{\varphi\varphi}/(g^2_{t\varphi} - g_{tt}\,g_{\varphi\varphi})]
   \right\} \;.
   \mbox{\raisebox{4ex}{}}
   \mbox{\raisebox{-4ex}{}}
\end{equation}

As our interest lies in analysing the `dragging' induced by the given 
spacetime, we shall consider only the orbit along which the 
{\em centrifugal acceleration is zero}. This as mentioned earlier is
possible in the stationary case as these orbits {\em do not} coincide with
unstable photon orbits as in the static case.

In such a situation one has for a particle in circular orbit on the 
equatorial plane $\theta = \pi/2$, the centrifugal acceleration along the
radial direction to be zero if
\begin{equation}
\label{def_R}
\left[
\frac{\partial_r g_{\varphi\varphi}}{g_{\varphi\varphi}} 
  - \frac{1}{2}
  \frac{\partial_r (g^2_{t\varphi} - g_{tt}\,g_{\varphi\varphi})}{
  (g^2_{t\varphi} - g_{tt}\,g_{\varphi\varphi})}
\right]_{\theta = \pi/2} = 0
  \mbox{\raisebox{4ex}{}}
  \mbox{\raisebox{-4ex}{}}
\end{equation}
For specified $g_{ij}$, this would give an algebraic equation, the real 
roots of which correspond to the location $R$ of the orbits with the required
condition. Evaluating the gravitational and Coriolis accelerations at this
location $(G)_{R,\pi/2}$ and $(Co)_{R,\pi/2}$ from 
(\ref{def_Gi}) and (\ref{def_Co}) one can get these two forces acting on the
particle as functions of the angular velocity parameter $\Omega$ and the
rotation parameter associated with the background spacetime. 

We then define the {\em `cumulative drag index'}
\begin{equation}
\label{def_calC}
{\cal C} = [(Co)_R - (G)_R]/[(Co)_R + (G)_R]
  \mbox{\raisebox{4ex}{}}
  \mbox{\raisebox{-4ex}{}}
\end{equation}
as the ratio of the relative drag acceleration to the total acceleration
acting on the particle as measured by the locally non rotating observer.

\section{Specific examples}
\subsection{Kerr geometry}
The spacetime metric outside a rotating black hole is, as 
well known, given by
\begin{equation}
\label{Kerr_metric}
ds^2 = -(1 - \frac{2 m r}{\Sigma})dt^2
  - \frac{4 m r a}{\Sigma}\sin^2\theta\,dt\,d\varphi
  + \frac{\Sigma}{\Delta} dr^2 + \Sigma\,d\theta^2 
  + \frac{B}{\Sigma}\sin^2\theta d\varphi^2
  \mbox{\raisebox{4ex}{}}
  \mbox{\raisebox{-4ex}{}}
\end{equation}
with 
$B = (r^2 + a^2)^2 - \Delta\,a^2\sin^2\theta$,
$\Delta = r^2 + a^2 - 2 m r$,
$\Sigma = r^2 + a^2\cos^2\theta$.

Using these $g_{ij}$ s in (\ref{def_Gi}) to (\ref{def_Cf}) the accelerations
may be obtained explicitly on the equatorial plane $\theta = \pi/2$ as
(see also Nayak and Vishveshwara \cite{NayVis})
\begin{eqnarray}
 & & \nonumber \\
(G)_r & = & -\frac{m (a^2\Delta + r^4 + a^2 r^2 - 2 m a^2 r)}{
  r \Delta(r^3 + a^2 r + 2 m a^2)}
  \label{GrKerr} \\
  & & \nonumber \\  
(Co)_r & = & \frac{2 m a (\Omega - \omega) (3 r^2 + a^2)}{
  r [1 - \Omega^2(r^2 + a^2) - (2 m/r) (1 - \Omega a)^2] 
  (r^3 + a^2 r + 2 m a^2)}
  \label{CorKerr} \\
   & & \nonumber 
\end{eqnarray}
and
\begin{equation}
\label{CfrKerr}
(Cf)_r = -\frac{(\Omega - \omega)^2
  [r^5 - 3 m r^4 + a^2 (r^3 - 3 m r^2 + 6 m^2 r - 2 m a^2)]}{
  r^2\Delta[1 - \Omega (r^2 + a^2) - (2 m/r) (1 - \Omega a)^2]}
  \mbox{\raisebox{4ex}{}}
  \mbox{\raisebox{-4ex}{}} 
\end{equation}
with 
$\omega = 2 m a/(r^3 + a^2 r + 2 m a^2)$.

\begin{figure}[p]
\centerline{\epsffile{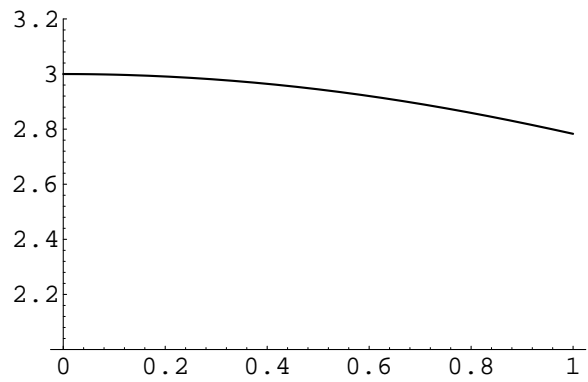}}
\caption{\label{fig1} Location of $R$ ($\uparrow$) as a function of $a$ 
($\rightarrow$).}
\vspace*{4cm}
\end{figure}

It is well known that when $(\Omega - \omega)$ the angular velocity of the
particle as measured by the Bardeen observer (LNRF) is zero, both Coriolis
and centrifugal force vanish identically.
On the other hand as seen from above, in this formulation of inertial 
forces, one can have the centrifugal force to be zero for 
$(\Omega - \omega) \neq 0$ if the algebraic expression
\begin{equation}
\label{R_Kerr}
r^5 - 3 m r^4 + a^2 (r^3 - 3 m r^2 + 6 m^2 r - 2 m a^2) = 0
  \mbox{\raisebox{4ex}{}}
  \mbox{\raisebox{-4ex}{}}
\end{equation} 
(Iyer and Prasanna 1992 \cite{IyePra}) has real positive roots.
It may be seen that for $0 < a < 1$, there are atmost three real roots of 
which at least one lies outside the event horizon. Fig.~(\ref{fig1}) shows
the location of this root as a function of $a$, the Kerr parameter.
Denoting this root by $R$, one can calculate $(G)_R$ and $(Co)_R$ from
(\ref{GrKerr}) and (\ref{CorKerr}) and finally obtain the cumulative drag 
index to be
\begin{equation}
\label{calCKerr}
{\cal C} = \frac{[2 m a (\Omega - \omega) ( 3 R^2 + a^2)]\,\Delta
  + m S\,(a^2\Delta + R^4 + a^2 R^2 - 2 m a^2 R)}{
  [2 m a (\Omega - \omega) (3 R^2 + a^2)]\,\Delta
  - m S\,(a^2\Delta + R^4 + a^2 R^2 - 2 m a^2 R)}
  \mbox{\raisebox{4ex}{}}
  \mbox{\raisebox{-4ex}{}}
\end{equation}
with 
$S = 1 - \Omega^2 (R^2 + a^2) - 2 m (1 - \Omega a)^2/R$.

\begin{figure}
\centerline{\epsffile{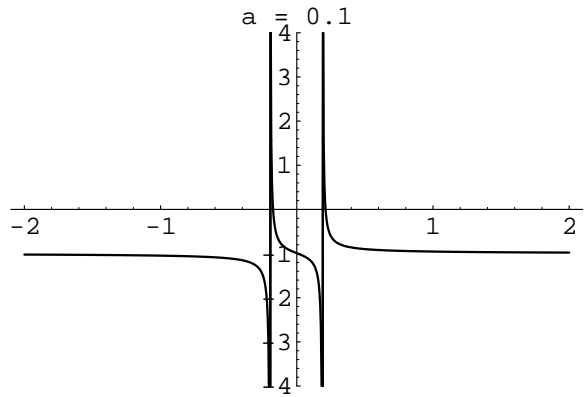}}
\centerline{\epsffile{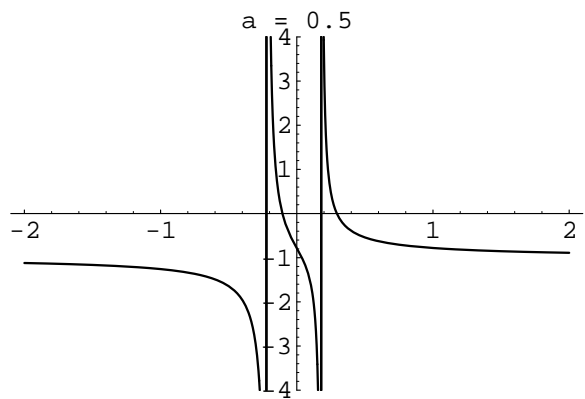}}
\centerline{\epsffile{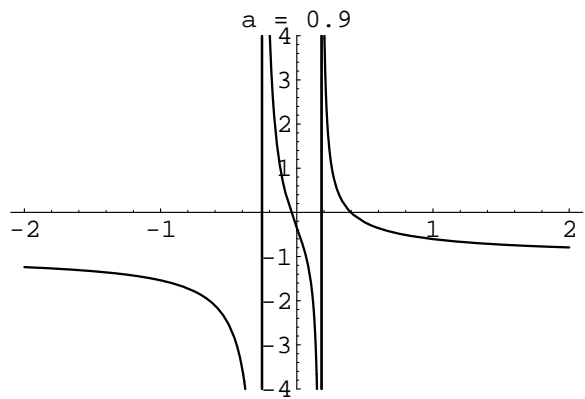}}
\caption{\label{fig2} Plots of ${\cal C}$ ($\uparrow$) as a function of 
$\Omega$ ($\rightarrow$) for different values of $a$.}
\end{figure}

Fig.~(\ref{fig2}) gives the plots of ${\cal C}$ as a 
function of $\Omega$ for different specific values of $a$. As seen from the
figures the function ${\cal C}$ has two zeros and two infinities as may
be expected from the fact that both the numerator and the denominator are
quadratic functions of $\Omega$.

${\cal C}$ is positive only for a very narrow range of values of 
$|\Omega|$ $(\ll 1)$ whereas it is negative for all other values of 
$\Omega$. Further, both the co rotating $(a\Omega > 0)$ and the counter
rotating $(a\Omega < 0)$ ones have exactly the same range of values of 
$\Omega$ for which ${\cal C}$ is positive, and this range increases with 
$a$. (Table \ref{tab1}).

Nayak and Vishveshwara (1996) have calculated the gyroscopic precession rate
$\tau_1$ in the Kerr geometry. Using their expression for $\theta = \pi/2$
plane, the precession rate is given by
\begin{equation}
\label{GypKerr}
\tau_1 = \frac{\Omega r^3 - 3 m \Omega (1 - \Omega a) r^2 
  + m a (1 - \Omega a)^2}{
  r^2 [-\Omega^2 r^3 + (1 - \Omega^2 a^2) r - 2 m (1 - \Omega a)^2]}
  \mbox{\raisebox{4ex}{}}
  \mbox{\raisebox{-4ex}{}}
\end{equation}
which when $a = 0$, is identically zero at $r = 3 m$, where the centrifugal
force also vanishes in Schwarzschild spacetime. On the other hand in the
present case for $r = R$, where $Cf = 0$, we have (after rescaling with $m$)
\begin{equation}
\label{GypCf0Kerr}
\tau_1 = \frac{(\alpha^3 + 3\alpha R^2)\Omega^2 
  + (R^3 - 3 R^2 - 2 \alpha^2)\Omega + \alpha}{
  -R^2 (R^3 + R \alpha^2 + 2 \alpha^2)\Omega^2 
  + 4 \alpha R^2 \Omega + (R^3 - 2 R^2)} \;.
  \mbox{\raisebox{4ex}{}}
  \mbox{\raisebox{-4ex}{}}
\end{equation}

\begin{figure}
\centerline{\epsffile{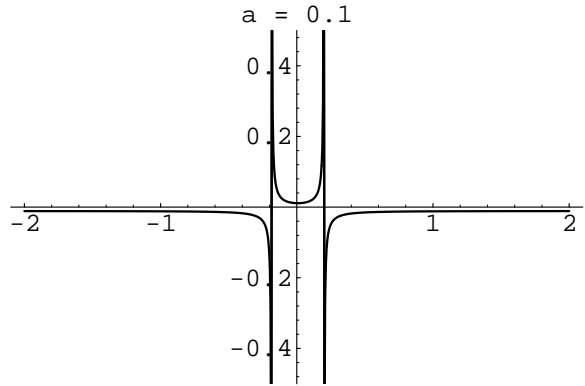}}
\centerline{\epsffile{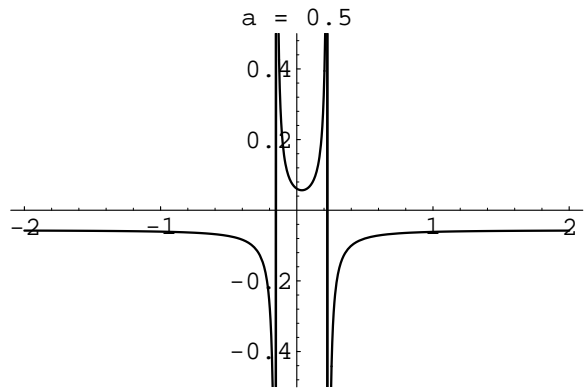}}
\centerline{\epsffile{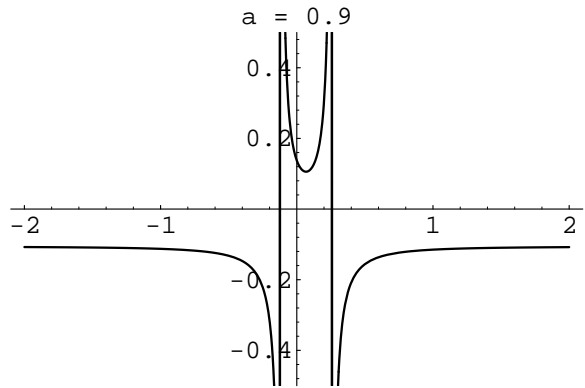}}
\caption{\label{fig4} Gyroscopic precession rate $\tau_1(R)$ 
as a function of $\Omega$ ($\rightarrow$).}
\end{figure}

\begin{figure}
\centerline{\epsffile{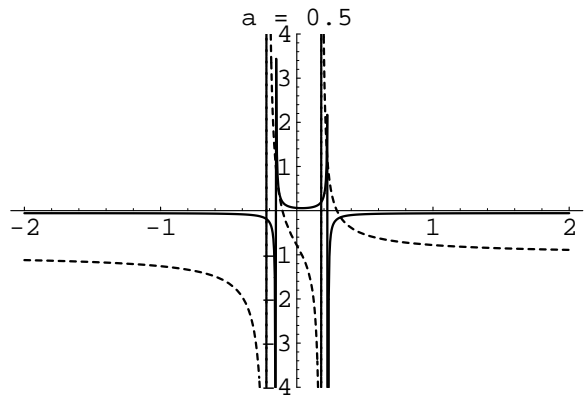}}
\centerline{\epsffile{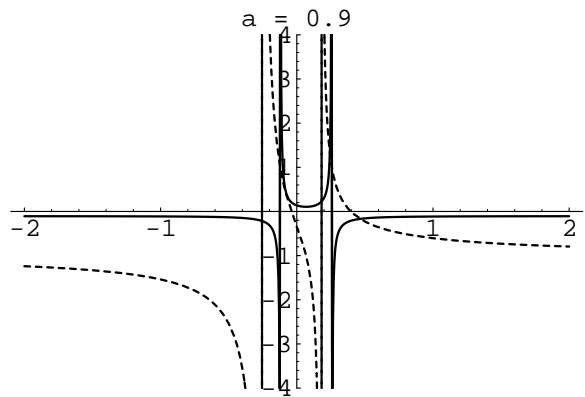}}
\caption{\label{fig5} ${\cal C}$ ($\cdots$) and $\tau_1(R)$ (---) as a  
function of $\Omega$ ($\rightarrow$).}
\end{figure}

As may be seen easily this would have real zeros if  and only if $R$, which
is also a function of $\alpha = a/m$ (eq.~\ref{R_Kerr}) satisfies the 
relation $R (R - 3)^2 > 4 \alpha^2$. On the other hand $\tau_1$ is infinite
for 
\[
\Omega_\pm = (2 \alpha \pm R \Delta^{1/2})/(R^3 + R \alpha^2 + 2 \alpha^2)
  \mbox{\raisebox{4ex}{}}
  \mbox{\raisebox{-4ex}{}}
\]
outside the event horizon ($\Delta = 0$) and further is positive in the
entire range
\[
[(2 \alpha - R \Delta^{1/2})/(R^3 + R \alpha^2 + 2 \alpha^2)] 
< \Omega < 
[(2 \alpha + R \Delta^{1/2})/(R^3 + R \alpha^2 + 2 \alpha^2)]
  \mbox{\raisebox{4ex}{}}
  \mbox{\raisebox{-4ex}{}}
\]
and negative outside this range (fig.~\ref{fig4}).
It is interesting to note (fig.~\ref{fig5}) that for values of
$\Omega$ where the index ${\cal C}$ is zero, $\tau_1$ is negative for 
co rotating particles and positive for counter rotating particles. 

\subsection{G\"odel spacetime}
One of the simplest homogeneous cosmological model is G\"odel's solution
of Einstein's equations which represents a pressure free, dusty universe 
having a non zero cosmological constant as given by the spacetime metric
(Hawking and Ellis 1972 \cite{HawEll})
\begin{equation}
\label{Goedel_metric}
ds^2 = 2 a^2\,\left(
   -dt^2 + dr^2 + dz^2 + f(r)\,d\varphi^2 + g(r)\, d\varphi\,dt\right)
  \mbox{\raisebox{4ex}{}}
  \mbox{\raisebox{-4ex}{}}
\end{equation}
with $a^2 = 1/8\pi\rho$, $f = \sinh^2 r - \sinh^4 r$, 
$g = 2\sqrt{2}\sinh^2 r$. $\rho$ being the matter density which should be
positive everywhere. Using these metric functions in the expressions 
(\ref{def_Gi}) to (\ref{def_Cf}) for the accelerations of a test particle
in circular orbit one finds:
\begin{eqnarray}
 & & \nonumber \\
(G)_r  & = & - g (2 f g^\prime - f^\prime g)/2 f (f + g^2) \\
 & & \nonumber \\
(Co)_r & = & -2 A^2 (\Omega - \omega) a^2 (g f^\prime - g^\prime f)/f \\
 & & \nonumber \\
(Cf)_r & = & -A^2 (\Omega - \omega)^2 2 a^2 
   [(2 g^2 + f) f^\prime - 2 g g^\prime f]/(g^2 + f) f \\
 & & \nonumber    
\end{eqnarray}
with 
$\omega = -g/f$, 
$A^2 = [2 a^2 (1 - 2 \Omega g - \Omega^2 f)]^{-1}$.
If we now consider the particular orbit on which the centrifugal force is
zero we get
\begin{equation}
[(2 g^2 + f) f^\prime - 2 g g^\prime f]_R = 0
  \mbox{\raisebox{4ex}{}}
  \mbox{\raisebox{-4ex}{}} 
\end{equation}
and using this in the other two we get
\begin{equation}
(G)_R = - f^\prime/2 f\;, \quad
(Co)_R = 4 a^2 A^2 (\Omega - \omega) f^\prime/2 g \;.
  \mbox{\raisebox{4ex}{}}
  \mbox{\raisebox{-4ex}{}}
\end{equation}
With these, the index ${\cal C}$ turns out to be
\begin{equation}
{\cal C} = (f\Omega + 2 g) (1 - g \Omega)/\Omega (f + 2 g^2 + \Omega f g)
  \mbox{\raisebox{4ex}{}}
  \mbox{\raisebox{-4ex}{}}
\end{equation}
which clearly shows the two zeros and the two infinities of  ${\cal C}$. 
Fig.~(\ref{fig6}) shows ${\cal C}$ as a function of $\Omega$ and again like
in the earlier case, ${\cal C}$ is positive only for a very small range of
values {\em viz.}, $-(f + 2 g^2)/f g < \Omega < -2g/f$ and 
$0 < \Omega < 1/g$, which again is the same for both co and counter 
rotating particles.

\section{Discussion}
It is indeed very significant that the `cumulative drag index' has the same
sign for both co and counter rotating particles in any given range of values
of the angular velocity parameter $\Omega$, irrespective of the spacetime
geometry. As the two examples sighted above are very typical for the 
discussion of `inertial frame dragging', the index could characterise the
asymptotic effects if there are any. In fact the G\"odel universe though is
pathological (being achronal) is really well suited to study pure geometric
effects as it is homogeneous and pressure free. The Kerr solution on the other
hand being the spacetime devoid of any matter distribution, is again free from
pressure gradient or electromagnetic effects. Thus for a particle in 
circular orbit in these spacetimes, the only forces acting on it are the
gravitational, Coriolis, and centrifugal. By considering the orbit where
the centrifugal force is zero, we have restricted futher the influence arising
from the spacetime surrounding the particle. If the positivity of the 
cumulative drag index is to show the influence of the distant universe
on the particle, then the fact that the influence on both co and counter
rotating particles exist to the same extent as measured by the locally non
rotating observer is of special interest.

\begin{figure}
\centerline{\epsffile{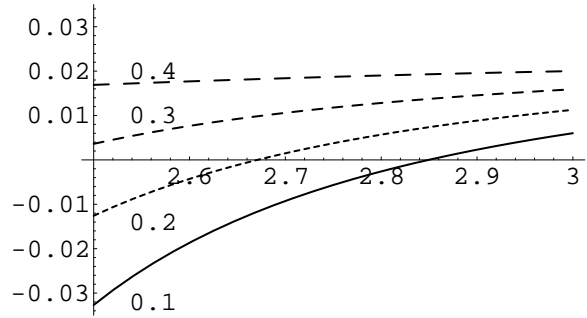}}
\caption{\label{fig7} $\tau_1(r)$ ($\uparrow$) as a function of 
$r$ ($\rightarrow$) for $\Omega = 0.1$ and different values of $a$ 
($0.1$ to $0.4$).}
\end{figure}

\begin{figure}
\centerline{\epsffile{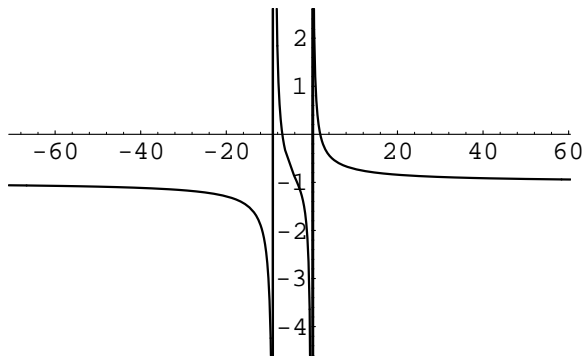}}
\caption{\label{fig6} ${\cal C}$ ($\uparrow$) for G\"odel universe, as a 
function of $\Omega$ ($\rightarrow$).}
\end{figure}

Table \ref{tab2} gives the locations of the zeros and infinities of the 
general function $\tau_1(r)$ (\ref{GypKerr}) for different values of $a$ for
$\Omega = 0.1$. The precession rate $\tau_1$ has no zeros for $a$
higher than a critical value for every $\Omega$, as is clear from
fig.~(\ref{fig7}). As observed by Nayak and Vishveshwara $\tau_1$ is zero
at a different location than the zeros of the centrifugal force and thus there 
always exists a non zero precession even when the cumulative drag index is 
zero. It is interesting to note that this precession corresponding to
${\cal C} = 0$ is {\em negative} for {\em co rotating} particles and
{\em positive} for {\em counter rotating} particles.

In the above we have restricted the analysis to the circular motion of the
particle in the equatorial plane. However, one might get more information
if one considers the behaviour of the drag index for other values of 
$\theta$, particularly to compare with the precession rate at poles and
the equator, of the gravitating sources. In fact the Lense--Thirring dragging
effect is also considered as a kind of gravimagnetic effect (Will 1995)
\cite{Wil}) by comparing the dragging of inertial frames to the influence of
the magnetic field
of a rotating electrical conductor. It would indeed be useful to consider the 
discussion of the `forces' to including the electromagnetic
fields and then look for an index when both centrifugal and Coriolis
forces are absent, thus having only gravitational and electromagnetic effects
in static geometry. Comparing the behaviour of such an index with ${\cal C}$,
one can perhaps get a better feeling about the local and global effects
associated with rotation.

\section*{Acknowledgement}
It is a great pleasure to thank Urs M. Schaudt for his help in getting the
plots as well as in preparing the ms in the \LaTeX\ mode. Discussions with 
Urs M. Schaudt, Herbert Pfister, Jiri Bicak,Jurgen Ehlers and Wolfgang Rindler,
who went through the manuscript, helped in clarifying many points and in the final
presentation. The work was done while the author was visiting the Institut 
f\"ur Theor. Phys., Universit\"at T\"ubingen,Germany, during the spring of 1996. 
He is grateful to the Alexander von Humboldt
Stiftung, Bonn, for the financial grant which made this visit posible.

\newpage
\begin{table}[p]
\caption{\label{tab1} Gives the $\Omega$ values for co (+) and counter (-)
rotating particles, for which ${\cal C}$ is zero (1) and infinity (2).}
\bigskip
\centerline{
\begin{tabular}{|l|l|l|l|l|l|l|}\hline
$a$ & $R$ & $\Omega_{1+}$ & $\Omega_{2+}$ &
$\Omega_{1-}$ & $\Omega_{2-}$ &
$(\Omega_1 - \Omega_2)$ \\ \hline\hline
0.1 & 2.9978 & 0.211269 & 0.189022 & -0.1742 & -0.1964 & 0.02224 \\ 
\hline
0.5 & 2.9445 & 0.29437 & 0.180094 & -0.10537 & -0.21965 & 0.11428 \\
\hline
0.9 & 2.8226 & 0.39668 & 0.17724 & -0.04084 & -0.26027 & 0.21944 \\
\hline
1.0 & 2.7830 & 0.42640 & 0.17722 & -0.02535 & -0.27452 & 0.24917 \\
\hline
\end{tabular}
}
\end{table}

\begin{table}[p]
\caption{\label{tab2} Gives the location of the event horizon ($r_e$) and the
location of the infinities and zero of $\tau_1(r)$.}
\bigskip
\centerline{
\begin{tabular}{|l|l|l|l|l|}\hline
$a$ & $r_e$ & $r_{1\infty}$ & $0$ & $r_{2\infty}$ \\ 
\hline\hline
0.1 & 1.99499 & 2.046606 & 2.84927 & 8.81822 \\
\hline
0.5 & 1.86603 & 1.87568 & $\;$-- & 8.91667 \\
\hline
0.9 & 1.43581 & 1.72113 & $\;$-- & 8.98669 \\
\hline
1.0 & 1.0 & 1.68466 & $\;$-- & 9.0 \\
\hline
\end{tabular}
}
\end{table}

\newpage

\newpage


\newpage

\newpage


\newpage

\newpage

\end{document}